# Polarity determination in ZnSe nanowires by HAADF STEM

**M. Den Hertog[1], M. Elouneg-Jamroz[1], E. Bellet-Amalric[2], S. Bounouar[2], C. Bougerol[1], R. André[1], Y. Genuist[1], J.P. Poizat[1], K. Kheng[2] and S. Tatarenko[1].**

CEA-CNRS group "Nanophysique et Semiconducteurs"

[1] Institut Néel, CNRS et Université Joseph Fourier, 25 rue des Martyrs, BP 166, 38042 Grenoble Cedex 9, France

[2] CEA-Grenoble INAC-SP2M, 17 rue des Martyrs, 38054 Grenoble Cedex 9, France

Martien.den-hertog@grenoble.cnrs.fr

**Abstract**. High angle annular dark field scanning transmission electron microscopy is used to analyze the polarity of ZnSe nanowires grown, by molecular beam epitaxy, on GaAs substrates. The experimental results are compared to simulated images in order to verify possible experimental artefacts. In this work we show that for this type of nano-objects, a residual tilt of the specimen below 15 mrad, away from the crystallographic zone axis does not impair the interpretation of the experimental images.

## 1. Introduction

ZnSe nanowires (NWs) are considered interesting structures for novel electronic and opto-electronic device functionalities. In ZnSe NWs both the hexagonal wurtzite (WZ) and cubic zinc blende (ZB) structure can occur [1]. These crystallographic structures for binary compounds are non-centrosymmetrical. Typically, along <111> in ZB and <0001> in WZ, one can define a polarity of the wire based on whether the II to VI (Zn to Se) bonds are aligned in the growth direction or opposite to it. It's important to investigate the polarity of such materials as it can affect their physical properties, in particular their optical properties. Indeed the crystal polarity is at the origin of electrical polarization along the growth direction: piezoelectric polarization in ZB or spontaneous polarization in WZ structures. In this paper we investigate the direction of the Zn to Se bond; referred to as the polarity, of ZnSe NWs epitaxially grown on GaAs substrates using High Angle Annular Dark Field Scanning Transmission Electron Microscopy (HAADF STEM). We do not study a possible effect of the direction of the ZnSe bond, the presence of polarization fields. HAADF STEM is an efficient technique to probe polarity along nanowires as it is sensitive to chemical contrast, namely sensitive to the atomic number Z. The experimental results are compared to simulated HAADF images to evaluate how the experimental conditions could affect the reliability of the polarity determination.

## 2. Experimental details

The NWs are grown by molecular beam epitaxy (MBE) using gold as a catalyst. Samples used in this study are all grown in a solid source MBE system. The source materials for the MBE system are elemental Zn (6N) and Se (6N). Commercial GaAs (111)B substrates are first deoxidized by annealing at 640°C under As pressure in a connected III-V MBE growth chamber. We then grow a 70 nm GaAs buffer layer. In order to avoid Ga incorporation in the NWs and to improve the surface quality and the epitaxial relation between the NWs and the substrate, a thin ZnSe buffer layer (about 50 nm thick) is grown at 280°C as described in [2]. Gold is deposited from an effusion cell on the ZnSe buffer layer, at room temperature, in a dedicated metal deposition chamber connected by UHV to the II-VI and III-V growth chambers. The nominal thickness of gold is on the order of 0.1 nm. The sample was heated at 500°C to cause dewetting of the gold layer and create nanocatalyst particles. ZnSe NWs are grown between 300°C and 450°C with a Zn:Se beam equivalent pressure (BEP) ratio of 1:4 and pressures in the $10^{-7}$ Torr range.

HAADF STEM images were obtained on a probe Cs corrected FEI TITAN TEM (300 kV). TEM samples were prepared by classical mechanical polishing followed by Ar ion milling using a Gatan PIPS. HAADF STEM multi slice simulations were realized using the code developed by Kirkland [3], using a 6x6 zinc blende (ZB) ZnSe unit cell with a sample thickness of 8 nm, corresponding to 20 (110) planes and comparable to the NW diameter (Fig. 1B). The input potential and wavefunction were sampled with 512 px and the final image or linescan was made on a length of 1.2 nm sampled with 64 px. The electron optical parameters of the Titan, that were included in the simulation, were 300 kV acceleration voltage, a spherical aberration of 10 µm, a convergence angle of a 15 mrad, and defocus of 0 nm. The defocus was not known exactly, however, according to simulation a small defocus (for instance 10 nm) decreases the contrast and the contrast difference between Zn and Se atoms but does not affect the visibility of the polarity. The contributions of the chromatic aberration are not included in these calculations. Images were calculated using 50 and 200 mrad inner and outer angles for the HAADF detector.

## 3. Results and Discussion

An SEM image of the ZnSe NW sample is shown in Fig. 1A. It can be observed that most of the NWs are vertical, parallel to the <111> normal of the substrate, i.e. very likely in epitaxial relation with the ZnSe buffer layers grown on the GaAs (111)B substrate. A few of them are inclined following the other <111> directions of the substrate. The typical diameter of the wires is 10 nm (Fig. 1B) and their length 150 nm.

An HAADF STEM image of a NW from this sample is shown in Fig. 1B. Many stacking faults are present along the growth direction giving rise to ZB and WZ regions, growing either [111]-oriented or [0001]-oriented. An averaged zoom of the NW of Fig. 1B is shown in Fig. 1C, where three atomic planes can be observed: the two lower planes have the WZ structure and the two upper planes have the ZB structure respectively, alternatively we could describe Fig. 1C by the ZB structure with a rotational twin or stacking fault between the bottom and the middle plane. The signal to noise ratio in this image is improved by averaging over several regions in the NW around and including the boxed region in Fig. 1B [4]. Thanks to the Cs probe correction of the Titan TEM, the spatial resolution enables us to clearly resolve the dumbbells in the crystal lattice and a sufficient signal to noise ratio allows determining the polarity of the NWs. On most dumbbells a brighter contrast is observed in the top part of the dumbbell. Since growth was carried out on GaAs (111)B, where As (Z=33), the heavier atom, is up in the dumbbell, we find correspondingly that in the NW the heavier atom, being Se (Z=34) is up in the dumbbell. Using an atomic model of the structure we can see that the Zn to Se bond lying in the image plane is pointing down the NW. It seems that irrespective of defects present along the NW, the polarity in the wire follows the polarity imposed by the substrate. In the following these results are compared to simulated HAADF STEM images to investigate possible experimental artifacts.

HAADF STEM experimental images have to be compared to simulations in order to evaluate how much this approach is sensitive to misalignment of the samples regarding polarity determination. Evaluation of the effects of a small specimen tilt on the image is necessary as it is difficult to align the specimen exactly on the crystallographic axis. The diffraction pattern (and ronchigram in STEM) is easily used to orient the specimen on a crystallographic axis, however small residual specimen tilts below <1 deg (17 mrad) are difficult to observe but can obviously influence the image contrast. In [5] the effect of a small tilt away from the zone axis orientation was studied using multislice simulation for the example of silicon on different zone axis orientations. It was shown that in the case of a specimen tilt on the order of 1°, a reduction by 2 of the high-resolution image contrast should be expected. More importantly it was shown that the effect of a small crystal tilt can cause an asymmetry in the contrast on dumbbells in a silicon crystal on the [011] zone axis [5,6]. In the simulation, tilting of the specimen is equivalent to shifting the probe function between slices (and is only valid for small tilts of no more than ~17 mrad). Shifting of the probe function between slices will cause asymmetric intensity on the two atoms in the dumbbell. Therefore extreme care should be taken in the determination of the polarity. In practice such small residual tilts can be measured by comparison of micro diffraction patterns with simulations [6].

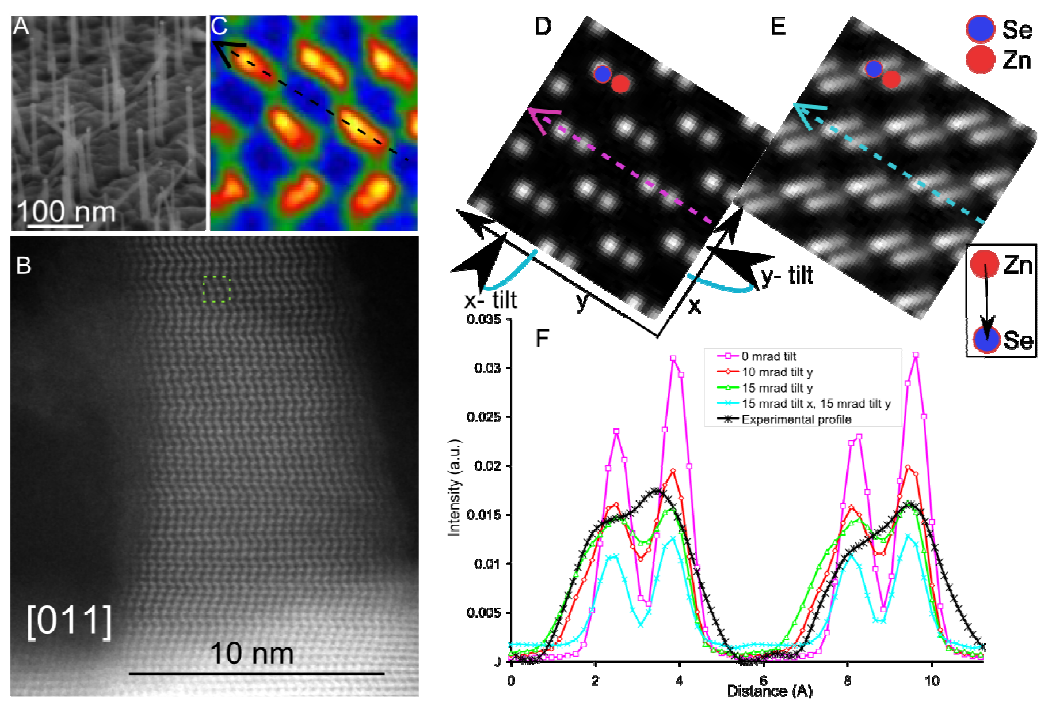

**Fig 1 Determination of the nanowire polarity and comparison with simulations.** (A) SEM image of NWs grown on GaAs(111)B substrate at 350°C (B) Experimental HAADF STEM image (filtered with a median filter using a 7 px square window) of the NW sample shown in (A). The wire grows in a <111> direction and contains many rotational twin defects/stacking faults on the growth plane. (C) Zoom averaged over several regions close to and including the squared region marked in (B), systematically a brighter contrast is observed on the upper atom in the dumbbell. (D) HAADF STEM simulation along the [011] direction for an 8 nm thick ZB ZnSe unit cell. The Zn and Se atoms are indicated respectively as well as the x and y tilting directions. (E) HAADF simulation with 15 mrad tilt in the x and y direction with respect to the previous simulation and phonon contributions at room temperature. The direction of the Zn to Se bond in the image plane is shown in the inset. (F) Simulated line profiles made along the arrows marked in (D,E) for different y and xy-tilt values off the [011] direction compared with a normalized experimental profile made in (C) along the black arrow, integrated over 0.05 nm.

In the case of epitaxially grown NWs, the specimen is aligned on the crystallographic axis of the substrate. However as the NWs are very thin and embedded in glue for sample preparation, they can be slightly bent and are therefore not necessarily oriented exactly the same as the substrate. Moreover, as the crystal volume of ZnSe NWs is extremely small, and these samples damage rapidly under the influence of the electron beam, micro diffraction measurements were not possible on these samples.

Instead we chose to simulate images and linescans in ZnSe with different specimen tilts up to 15 mrad to see if such tilts would influence the contrast on Zn and Se atoms.

Simulated HAADF STEM images are shown of ZB ZnSe on the [011] axis without specimen tilt (Fig. 1D) and with 15 mrad tilt in the x and y direction (Fig. 1E). The polarity is Se, meaning that the (111) plane is terminated Se. Clearly the polarity remains visible (meaning that the brightest contrast is observed on the heavier atom being Se) even at relatively important specimen tilts for this sample thickness. In Fig. 1F line scans are shown made along the dumbbells (dotted arrows Fig 1D,E) at different x and y tilts. It should be noticed that both x or y tilts weaken the intensity on the dumbbell but a tilt in the x direction preserves the intensity ratio of the two atoms while a tilt in the y direction lowers this ratio. We evidenced that the visibility of the polarity in the case of ZnSe NWs is rather independent of the specimen tilt because the specimen is thin, only 8 nm as compared to 25 and 35 nm considered in [5,6]. At our specimen thickness the difference in atomic number ($\Delta Z=4$) is sufficient to observe the polarity even if a specimen tilt close to 1° is present.

For comparison an experimental profile is shown obtained in Fig 1C. The experimental line profile was qualitatively normalized to allow comparison with the simulated profiles. The minimum intensity in the profile was set to zero and the maximum intensity was set to a value that allowed the best superposition of simulated and experimental profiles. The comparison of simulated and experimental profiles indicates that indeed residual specimen tilt is present. Furthermore the experimental profile is broader than the simulated profiles even at high tilt. Probably a broadening of the intensity is due to the amorphous material present around the NW, mainly glue used for the specimen preparation. This amorphous shell will cause some background noise in the image and loss of spatial resolution, but will not affect relative intensities on atomic columns since the amorphous shell is relatively homogeneous. Also mechanical vibrations can broaden the signal.

## 4. Conclusion

We have shown that the polarity in ZnSe Nws can be determined using HAADF STEM. In the case of epitaxially grown NWs the polarity is Se, as imposed by the GaAs starting substrate. Comparison with simulated images shows that even if a relatively large residual tilt of the specimen away from the crystallographic zone axis would be present, the polarity remains visible.

## 5. Acknowledgements

This work is supported by the French National Research Agency (ANR) through Nanoscience and Nanotechnology Program (Project BONAFO n°ANR-08-NANO-031-01) who provided a research fellowship for MdH. MEJ acknowledges financial support from the Nanosciences Foundation (RTRA). We acknowledge Jean-Luc Rouviere for use of the script to average over several regions in the STEM image.

**References**

[1] Xia D. et al, *Chin. Phys. Lett*. **23** (2006) 1317.
[2] Robin I.C., Aichele T., Bougerol C., André R et al., *Nanotechnology* **18** (2007) 265701.
[3] Kirkland E. J., Roane R. F. and Silcox J., *Utramicroscopy* **23** (1987) 77.
[4] Rouviere J.L., Bougerol C. et al, *Appl. Phys. Lett*. **92** (2008) 201904
[5] Maccagnano-Zacher S.E., Mkhoyan K.A., Kirkland E.J. et al., *Ultramicroscopy* **108** (2008) 718-726.
[6] Yamazaki T., Kawasaki M., Watanabe K., Hashimoto I., Shiojiri M., *Ultramiscroscopy* **92** (2002) 181-189.